\documentclass[11pt]{article}
\newcommand{\equal}{\!\!\!&=&\!\!\!}
\begin{document}
\abovedisplayshortskip 12pt
\belowdisplayshortskip 12pt
\abovedisplayskip 12pt
\belowdisplayskip 12pt

\title{{\bf A Lax Equation for the Non-Linear Sigma Model}}
\author{J. C. Brunelli$^{a}$, A. Constandache$^{b}$ and Ashok
Das$^{b}$  \\
\\
$^{a}$ Departamento de F\'\i sica, CFM\\
Universidade Federal de Santa Catarina\\
Campus Universit\'{a}rio, Trindade, C.P. 476\\
CEP 88040-900\\
Florian\'{o}polis, SC\\
Brazil\\
\\
$^{b}$ Department of Physics and Astronomy,\\
University of Rochester,\\
Rochester, NY 14627-0171\\
USA.}
\date{}
\maketitle

\begin{center}
{ \bf Abstract}
\end{center}

We propose a Lax equation for the non-linear sigma model which
leads directly to the conserved local charges of the system. We show
that the system has two infinite sets of such conserved charges
following from the Lax equation, much like
dispersionless systems. We show that the system has two Hamiltonian
structures which are compatible so that it is truly a bi-Hamiltonian
system. However, the two Hamiltonian structures act on the two distinct
sets of charges to give the dynamical equations, which is quite
distinct from the behavior in conventional integrable systems. We
construct  two
recursion operators which connect the conserved charges within a given
set as well as between the two sets. We show explicitly
that the conserved charges are in involution with respect to either of
the Hamiltonian structures thereby proving complete
integrability of the system. Various other interesting features are
also  discussed.

\newpage

\section{Introduction:}

The non-linear sigma model has been studied from various points of
view~\cite{abdallas}. In the context of $1+1$ dimensional
integrable systems, it is one of the most enigmatic models. The
equations for the principal chiral model, for example, take the
forms
\begin{eqnarray}
\partial_{\mu}J^{\mu,\,a} \equal  0\;,\nonumber\\
\noalign{\vspace{6pt}}
\partial_{\mu}J_{\nu}^{a} - \partial_{\nu}J_{\mu}^{a} -
f^{abc}J_{\mu}^{b}J_{\nu}^{c} \equal  0\;,\label{sigmamodelequations}
\end{eqnarray}
where $\mu = 0,1$ and $J_{\mu}^{a}$ belongs to a simple Lie
algebra, with $a$ taking values in the adjoint
representation. Explicitly, the two equations can be written as
\begin{eqnarray}
\partial_{t}J_{0}^{a} & = & \partial_{x}J_{1}^{a}\nonumber\\
\partial_{t}J_{1}^{a} & = & \partial_{x}J_{0}^{a} + f^{abc}
J_{0}^{b}J_{1}^{c}
\end{eqnarray}
It is well known that the equations (\ref{sigmamodelequations}) can
be combined into a single zero curvature condition by introducing
a one parameter family of connections that depend on a constant
spectral parameter~\cite{zerocur}. The zero curvature
representation is quite useful and leads directly to the infinite
number of conserved non-local charges of the
system~\cite{nonlocalcharges}. While an infinity of conserved
charges is a hallmark of integrable systems, these non-local
charges are not in involution with respect to the Hamiltonian
structure of the system~\cite{nonlocalalgebra}. Rather, they
satisfy a Yangian algebra~\cite{yangian}. While classical
integrability of the non-linear sigma model is known from
alternate arguments~\cite{inversescat,gurses} (see \cite{quantum}
for quantum integrability), a conventional proof of integrability
through the existence of an infinite number of conserved charges
in involution is lacking so far. Furthermore, it is not known
whether the system is bi-Hamiltonian, as is the case with most
integrable systems~\cite{bihamiltonian,das1}.

There have been various attempts \cite{localcharges} to construct
conserved  local charges for the system
(\ref{sigmamodelequations}), but a systematic study through a Lax
representation is lacking so far. In this letter, we will provide
a Lax equation for the non-linear sigma model equations
(\ref{sigmamodelequations}). Although the equations
(\ref{sigmamodelequations}) cannot be characterized as
dispersionless, they share many features common to a
dispersionless model~\cite{dispersionless}. In section 2, we
discuss briefly the two component Riemann equation.  This is a
dispersionless system sharing some properties with the non-linear
sigma model. In section 3, drawing from our experience in the Toda
lattice~\cite{das1,okubo}, we present a Lax representation for
(\ref{sigmamodelequations}). Even though the Lax equation for the
system is not that of a dispersionless system, it immediately
leads to two infinite sets of local conserved charges, a common
feature of dispersionless systems~\cite{nutku,das}. We point out
various properties of these conserved charges that are quite
useful in proving the integrability of the system. We also
present the two Hamiltonian structures which are compatible and
make the  system bi-Hamiltonian. In section 4, we construct two
recursion operators for the system which connect charges within
any given set as well  as between the two sets. We also show
explicitly that the two sets of charges are in involution with
respect to either of the Hamiltonian structures  thereby proving
the integrability of the system. In section 5, we summarize our
results and discuss some other interesting aspects of this system.

\section{Matrix Riemann equation:}

Before discussing the Lax equation for the non-linear sigma
model, we will briefly discuss, in this section, the two
component Riemann equation. We recall that the Riemann equation
\begin{equation}
u_{t} = uu_{x}\;,
\end{equation}
is an integrable dispersionless system~\cite{nutku,brunelli}
which belongs to the class of equations of hydrodynamic type. Even
though the non-linear sigma model is not a dispersionless system,
it shares some nice properties with the two component Riemann
equation. Let us define a $2\times 2$ matrix
\begin{equation}
U = \left(\begin{array}{cc}
u & v\\
v & u
\end{array}\right) = u I + v \sigma_{1}\;,\label{U}
\end{equation}
where $u,v$ are the two dynamical variables of the system and
$\sigma_{1}$ is  the familiar Pauli matrix ($I$ denotes the identity
matrix.). Then, the generalized
two component Riemann equation takes the form
\begin{equation}
U_{t} = UU_{x}\;.\label{Riemann}
\end{equation}

Let us next consider a matrix Lax function of the type
\begin{equation}
L = p^{2} I + U\;,\label{Lax}
\end{equation}
where $p$ denotes the momentum in the phase space. Then, from the
special structure of $U$ in (\ref{U}), it is clear that
\begin{equation}
\left[\sigma_{1},U\right] = 0\;,\qquad \left[\sigma_{1},L\right] =
0\;.
\end{equation}
As a result, there is an ambiguity in taking the square root of the
Lax function, namely,
\begin{equation}
L_{1}^{\frac{1}{2}} = p\,I + \frac{1}{2} U p^{-1}- \frac{1}{8}
U^{2}p^{-3} + \cdots\;.
\end{equation}
Furthermore, a second square root has the form
\begin{equation}
L_{2}^{\frac{1}{2}} = \sigma_{1} L_{1}^{\frac{1}{2}}\;.
\end{equation}
We note that the square root (either one) of the Lax function commutes
with $\sigma_{1}$, as does $L$. Both the square roots lead to
consistent  Lax equations~\cite{das} as
\begin{eqnarray}
\frac{\partial L}{\partial t_{n}} \equal
\frac{(2n)!!}{(2n+1)!!}\;\left\{(L_{1})^{\frac{2n+1}{2}}_{\geq
1},L\right\}\;,\nonumber\\
\noalign{\vspace{6pt}}
\frac{\partial L}{\partial t_{n}} \equal
\frac{(2n)!!}{(2n+1)!!}\;\left\{(L_{2})^{\frac{2n+1}{2}}_{\geq 1},
L\right\}\;,
\end{eqnarray}
where $n=0,1,2,\cdots$ and which lead to the hierarchy of Riemann
equations  of the form
\begin{eqnarray}
U_{t} \equal U^{n}U_{x}\;,\label{Riemann1}\\
\noalign{\vspace{6pt}}
U_{t} \equal \sigma_{1}U^{n}U_{x}\;,\label{Riemann2}
\end{eqnarray}
respectively. In the second equation the effect of  $\sigma_{1}$, on
the right hand side,
is to interchange $u\leftrightarrow v$, and eq. (\ref{Riemann2}) can
be  thought of
as the ``elastic medium'' equations~\cite{nutku}  associated with the
hierarchy of Riemann
equations (\ref{Riemann1}). It is worth noting here that, although
we have written the equations as non-standard equations, the same
equations also arise from the standard Lax equations~\cite{alex}.

The conserved charges of the system can now be easily obtained from
the Lax function. In fact, there are two infinite sets of conserved
charges which arise as~\cite{das}
\begin{eqnarray}
H_{n} \equal \int dx\;h_{n} = \frac{(2n)!!}{(2n-1)!!}\,{\rm
Tr}\,L^{\frac{2n-1}{2}}  =  \frac{(2n)!!}{(2n-1)!!}\,\int
dx\,{\rm tr}\,{\rm Res}\,L^{\frac{2n-1}{2}}= \int dx\,{\rm
tr}\,U^{n}\;, \nonumber\\
\noalign{\vspace{6pt}} \tilde{H}_{n} \equal \int
dx\,\tilde{h}_{n} = \frac{(2n)!!}{(2n-1)!!}\, {\rm
Tr}\,\sigma_{1} L^{\frac{2n-1}{2}} = \frac{(2n)!!}{(2n-1)!!}\,
\int dx\,{\rm tr}\,{\rm Res}\,\sigma_{1} L^{\frac{2n-1}{2}} =
\int dx\,{\rm tr}\,\sigma_{1}U^{n}\;,\nonumber\\\label{charges}
\end{eqnarray}
where ``tr'' denotes the trace over matrix indices, ``Res'' (Residue)
stands for the coefficient of the $p^{-1}$ term in the expansion and
$n=1,2,3,\dots$.
(Alternatively, we can
think of the two sets of charges as the traces of powers of $L_{1}$ and
$L_{2}$ respectively.) Explicitly, the two sets of conserved densities
have the closed forms
\begin{eqnarray}
h_{n} \equal 2 \sum_{k=0}^{\left[\frac{n}{2}\right]}
\left(\begin{array}{c}
n\\
2k
\end{array}\right)\,
u^{n-2k}v^{2k}\;,\nonumber\\
\noalign{\vspace{6pt}}
\tilde{h}_{n} \equal  2\sum_{k=1}^{\left[\frac{n+1}{2}\right]}
\left(\begin{array}{c}
n\\
2k-1
\end{array}\right)\, u^{n-2k+1}
v^{2k-1}.\label{charges1}
\end{eqnarray}

The conserved quantities are, of course, defined up to overall
normalization constants. Rescaling
\begin{equation}
(H_{n},\tilde{H}_{n}) \rightarrow
\frac{1}{n(n-1)}\,(H_{n},\tilde{H}_{n}),\qquad n>1,
\end{equation}
we see that the equations (\ref{Riemann1}) and (\ref{Riemann2}) can be
written in the Hamiltonian form respectively as
\begin{eqnarray}
U_{t} & = & \partial\,\frac{\delta H_{n+2}}{\delta U}\;,\nonumber\\
\noalign{\vskip 6pt}%
U_{t} & = & \partial\,\frac{\delta \tilde{H}_{n+2}}{\delta U}\;.
\end{eqnarray}
The conserved densities of the system are easily seen to satisfy the
usual  relations
for the two component hydrodynamic systems~\cite{nutku}, namely,
\begin{equation}
\frac{\partial^{2}h_{n}}{\partial u\,\partial u} =
\frac{\partial^{2}h_{n}}{\partial v\,\partial v}\;,\qquad
\frac{\partial^{2}\tilde{h}_{n}}{\partial u\,\partial u} =
\frac{\partial^{2}\tilde{h}_{n}}{\partial v\,\partial v}\;.\label{Relation}
\end{equation}
This system of equations is integrable. However, we will not get into
the details of this. Our interest has been only to discuss it briefly so
that we can bring out the similarities of this system with the
non-linear sigma model.

\section{Lax equation for the non-linear sigma model:}

The non-linear sigma model is not exactly a dispersionless
system. Consequently, we do not expect a Lax function in the phase
space to describe such a system. We find that, like the Toda lattice,
the non-linear sigma model does not have a scalar Lax
description, but can be given a Lax description much like the Toda
lattice~\cite{das1,okubo} in the following manner.

Let us consider the symmetric $2\times 2$ (multiplicative) Lax
operator of the form
\begin{eqnarray}
S^{ab} \equal \left(\begin{array}{cc}
J_{0}^{a}J_{0}^{b}+J_{1}^{a}J_{1}^{b} &
J_{0}^{a}J_{1}^{b}+J_{1}^{a}J_{0}^{b}\\
\noalign{\vspace{6pt}}
J_{0}^{a}J_{1}^{b}+J_{1}^{a}J_{0}^{b} &
J_{0}^{a}J_{0}^{b}+J_{1}^{a}J_{1}^{b}
\end{array}\right)\nonumber\\
\noalign{\vspace{6pt}}
 \equal (J_{0}^{a}J_{0}^{b}+J_{1}^{a}J_{1}^{b})\, I +
 (J_{0}^{a}J_{1}^{b}+J_{1}^{a}J_{0}^{b})\,\sigma_{1}\nonumber\\
 \noalign{\vspace{6pt}}
 \equal L^{a}L^{b}\;,\label{s}
\end{eqnarray}
where
\begin{equation}
L^{a} = J_{0}^{a} I + J_{1}^{a} \sigma_{1} =
\left(\begin{array}{cc}
J_{0}^{a} & J_{1}^{a}\\\noalign{\vspace{6pt}}
J_{1}^{a} & J_{0}^{a}
\end{array}\right)\;.
\end{equation}
Let us also introduce an anti-symmetric $2\times 2$ matrix operator
\begin{equation}
U^{ab} = \left(\delta^{ab} \partial + f^{abc} J_{1}^{c}\right) \sigma_{1} =
\left(\begin{array}{cc}
0 & \delta^{ab}\partial + f^{abc}J_{1}^{c}\\\noalign{\vspace{6pt}}
\delta^{ab}\partial + f^{abc}J_{1}^{c} & 0
\end{array}\right)\;.
\end{equation}
Then, it is easy to check that the Lax equation of the form of the Toda
lattice, namely,
\begin{equation}
\frac{\partial S^{ab}}{\partial t} = - \left[S,U\right]^{ab}\equiv
- \left(S^{ac}U^{cb}-U^{ac}S^{cb}\right)\;,\label{lax}
\end{equation}
leads to the equations for the non-linear sigma model
(\ref{sigmamodelequations}). ($U^{ab}$, here, is an operator and
should not be confused with the matrix $U$ in (\ref{U}) of the last
section, which
describes the dynamical variables of the system much like $S^{ab}$.)
Parenthetically, we would like to
remark here that the correspondence with the Toda lattice, however, is
not quite complete as the operator $U^{ab}$ cannot be obtained as
the Fr\'echet derivative of the equations for the sigma model
\cite{das1,okubo}.
Just like the two component Riemann equation of the last section,
it is clear that
\begin{equation}
\left[\sigma_{1},L^{a}\right] = 0\;,\qquad
\left[\sigma_{1},S^{ab}\right] = 0\;,\qquad
\left[\sigma_{1},U^{ab}\right] = 0\;,\label{symmetry}
\end{equation}
which defines an analogous symmetry in the non-linear sigma model.

Equation (\ref{lax}) defines the Lax equation for the non-linear
sigma model. As in the Toda lattice, it follows that the
conserved densities of the system can be obtained as traces of
powers of the Lax operator $S^{ab}$~\cite{das1,okubo}. In fact, because
of the symmetry (\ref{symmetry}) in the system, it is trivial to
check that there are two infinite sets of local conserved
densities in the system, namely,
\begin{equation}
h_{n} = \frac{1}{4n}\,{\rm Tr}\,S^{n}\;,\qquad \tilde{h}_{n} =
\frac{1}{4n}\,{\rm Tr}\,\sigma_{1}
S^{n}\;.
\end{equation}
Here, ``Tr'' denotes trace over the matrix indices as well as over
the internal indices. Using the definition (\ref{s}) as well as the
cyclicity property of the trace, it is easy to see that we can write
the conserved densities in the simpler form
\begin{equation}
h_{n} = \frac{1}{4n}\,{\rm tr}\, (L^{2})^{n}\;,\qquad \tilde{h}_{n} =
\frac{1}{4n}\, {\rm tr}\,
\sigma_{1} (L^{2})^{n}\;,\label{conserveddensity}
\end{equation}
where
\begin{equation}
L^{2} = L^{a}L^{a} = (J_{0}^{2}+J_{1}^{2}) I + 2(J_{0}J_{1})
\sigma_{1}\;,\label{lsquared}
\end{equation}
and ``tr'' denotes, as before, trace over matrix indices. In
(\ref{lsquared}) the internal indices of the terms inside the
parenthesis are summed. The similarity of these with the conserved densities
of the two component Riemann equation in (\ref{charges}) is remarkable
(we have chosen a particular normalization for simplicity). The first
few conserved densities for the non-linear sigma model have the
explicit forms,
\begin{eqnarray}
\begin{array}{l}
h_{1}=\displaystyle\frac{1}{2}\,(J_{0}^2+J_{1}^2)\;,\\
\noalign{\vspace{10pt}}
h_{2}=\displaystyle\frac{1}{4}\left(J_{0}^2+J_{1}^2\right)^{2}+
\left(J_{0}J_{1}\right)^{2}\;,\\
\noalign{\vspace{10pt}}
h_{3}=\displaystyle\frac{1}{6}\left(J_{0}^2+J_{1}^2\right)^{3}+
{2}\left(J_{0}^2+J_{1}^{2}\right)
\left(J_{0}J_{1}\right)^{2}\;,\\
\quad\,\,\,\vdots
\end{array}
&
\begin{array}{l}
\tilde h_{1}=\displaystyle (J_{0}J_{1})\;,\\
\noalign{\vspace{10pt}}
\tilde h_{2}=\displaystyle \left(J_{0}^2+J_{1}^{2}\right)
\left(J_{0}J_{1}\right)\;,\\
\noalign{\vspace{10pt}}
\tilde h_{3}=\displaystyle \left(J_{0}^2+J_{1}^{2}\right)^{2}
\left(J_{0}J_{1}\right)+\frac{4}{3}(J_{0}J_{1})^{3}\;.\\
\quad\,\,\,\vdots
\end{array}
\end{eqnarray}
We see that the local conserved densities in
(\ref{conserveddensity}) do not involve spatial derivatives, much like
the conserved densities of dispersionless systems. It is also  clear
now that the two sets of conserved
charges can be mapped into those of the two component Riemann equation
(\ref{charges}) or (\ref{charges1}) (up to normalization constants)
with the identifications
\begin{equation}
u = J_{0}^{2}+J_{1}^{2}\;,\qquad v = 2J_{0}J_{1}\;.
\end{equation}
With this mapping, even the equations for the sigma model
(\ref{sigmamodelequations}) go over to those of the two component
Riemann equation (\ref{Riemann2}) for $n=0$. However, such a map is
not one  to
one and, therefore, is not very useful. Nonetheless, it brings out an
interesting relation between a dispersionless system and the
non-linear sigma model.

As in hydrodynamic systems, in the case of the non-linear sigma model,
it can be shown that any local conserved density, $Q$ (without spatial
derivatives), must satisfy
\begin{equation}
\frac{\partial^{2}Q}{\partial J_{0}^{a}\,\partial J_{0}^{b}} =
\frac{\partial^{2}Q}{\partial J_{1}^{a}\,\partial
J_{1}^{b}}\;,
\end{equation}
which is the analogue of (\ref{Relation}). The conserved densities in
(\ref{conserveddensity}) can be seen to satisfy this. However, from
the form of the conserved densities in (\ref{conserveddensity}), it
can be checked that they satisfy additional relations which can be
written collectively as
\begin{eqnarray*}
\begin{array}{l}\displaystyle
\frac{\partial^{2}h_{n}}{\partial J_{0}^{a}\,\partial J_{0}^{b}} =
\frac{\partial^{2}h_{n}}{\partial J_{1}^{a}\,\partial
J_{1}^{b}}\;,\\
\noalign{\vspace{10pt}}\displaystyle
\frac{\partial^{2}h_{n}}{\partial
J_{0}^{a}\,\partial J_{1}^{b}} = \frac{\partial^{2}h_{n}}{\partial
J_{0}^{b}\,\partial
J_{1}^{a}}\;,\\
\noalign{\vspace{10pt}}\displaystyle
J_{0}^{a}\frac{\partial h_{n}}{\partial J_{0}^{a}} +
J_{1}^{a}\frac{\partial h_{n}}{\partial J_{1}^{a}}  =  2n
h_{n}\;,\\
\noalign{\vspace{10pt}}\displaystyle
J_{1}^{a}\frac{\partial h_{n}}{\partial J_{0}^{a}} +
J_{0}^{a}\frac{\partial h_{n}}{\partial J_{1}^{a}} =  2n
\tilde{h}_{n}\;,\\
\noalign{\vspace{10pt}}\displaystyle
f^{abc}\left(J_{0}^{b}\frac{\partial h_{n}}{\partial
J_{0}^{c}}+J_{1}^{b}\frac{\partial h_{n}}{\partial
J_{1}^{c}}\right) =  0\;,\\
\noalign{\vspace{10pt}}\displaystyle
f^{abc}\left(J_{0}^{b}\frac{\partial h_{n}}{\partial
J_{1}^{c}}+J_{1}^{b}\frac{\partial h_{n}}{\partial
J_{0}^{c}}\right) =  0\;,
\displaystyle
\end{array}&&
\begin{array}{l}\displaystyle
\frac{\partial^{2}\tilde h_{n}}{\partial J_{0}^{a}\,\partial J_{0}^{b}} =
\frac{\partial^{2}\tilde h_{n}}{\partial J_{1}^{a}\,\partial
J_{1}^{b}}\;,\\
\noalign{\vspace{7pt}}\displaystyle
\frac{\partial^{2}\tilde h_{n}}{\partial
J_{0}^{a}\,\partial J_{1}^{b}} = \frac{\partial^{2}\tilde h_{n}}{\partial
J_{0}^{b}\,\partial
J_{1}^{a}}\;,\\
\noalign{\vspace{7pt}}\displaystyle
J_{0}^{a}\frac{\partial\tilde h_{n}}{\partial J_{0}^{a}} +
J_{1}^{a}\frac{\partial\tilde h_{n}}{\partial J_{1}^{a}}  =  2n
\tilde h_{n}\;,\\
\noalign{\vspace{7pt}}\displaystyle
J_{1}^{a}\frac{\partial\tilde h_{n}}{\partial J_{0}^{a}} +
J_{0}^{a}\frac{\partial\tilde h_{n}}{\partial J_{1}^{a}} =  2n
{h}_{n}\;,\\
\noalign{\vspace{7pt}}\displaystyle
f^{abc}\left(J_{0}^{b}\frac{\partial\tilde h_{n}}{\partial
J_{0}^{c}}+J_{1}^{b}\frac{\partial\tilde h_{n}}{\partial
J_{1}^{c}}\right) =  0\;,\\
\noalign{\vspace{7pt}}\displaystyle
f^{abc}\left(J_{0}^{b}\frac{\partial\tilde h_{n}}{\partial
J_{1}^{c}}+J_{1}^{b}\frac{\partial\tilde h_{n}}{\partial
J_{0}^{c}}\right) =  0\;,
\end{array}
\end{eqnarray*}
\begin{equation}
\left(\begin{array}{c}\displaystyle
\frac{\partial{h}_{n}}{\partial J_{0}^{a}}\\
\noalign{\vspace{6pt}}\displaystyle
\frac{\partial{h}_{n}}{\partial J_{1}^{a}}
\end{array}\right) = \sigma_{1}\left(\begin{array}{c}\displaystyle
\frac{\partial\tilde h_{n}}{\partial J_{0}^{a}}\\
\noalign{\vspace{5pt}}\displaystyle
\frac{\partial\tilde h_{n}}{\partial J_{1}^{a}}
\end{array}\right)\;.
\label{extrarelations}
\end{equation}
These relations are quite useful in proving the complete integrability
of the system as well as various other interesting properties, as we
will see.

Once the two sets of local conserved charges have been obtained, it is
straightforward to show that the non-linear sigma model has two
Hamiltonian structures. Defining
\begin{equation}
{\cal D}_{1}^{ab} = \left(\begin{array}{cc}
0 & \delta^{ab} \partial\\\noalign{\vspace{6pt}}
\delta^{ab} \partial & -f^{abc}J_{0}^{c}
\end{array}\right)\;,\qquad {\cal D}_{2}^{ab} = \left(\begin{array}{cc}
\delta^{ab} \partial & 0\\\noalign{\vspace{6pt}}
0 & \delta^{ab} \partial + f^{abc} J_{1}^{c}
\end{array}\right)\;.\label{hamiltonianstructure}
\end{equation}
it is easy to see that the non-linear sigma model equations,
(\ref{sigmamodelequations}), can be written in the Hamiltonian form as
\begin{equation}
\left(\begin{array}{c}
J_{0,t}^{a}\\\noalign{\vspace{15pt}}
J_{1,t}^{a}
\end{array}\right) = {\cal D}_{1}^{ab} \left(\begin{array}{c}
\displaystyle\frac{\delta H_{1}}{\delta J_{0}^{b}}\\\noalign{\vspace{6pt}}
\displaystyle\frac{\delta H_{1}}{\delta J_{1}^{b}}
\end{array}\right) = {\cal D}_{2}^{ab} \left(\begin{array}{c}
\displaystyle\frac{\delta \tilde{H}_{1}}{\delta
J_{0}^{b}}\\\noalign{\vspace{6pt}}
\displaystyle\frac{\delta \tilde{H}_{1}}{\delta J_{1}^{b}}
\end{array}\right)\;.
\end{equation}
In fact, using the properties in (\ref{extrarelations}), it is easy to
show that all the higher order equations in the hierarchy can also be
written as
\begin{equation}
\left(\begin{array}{c}
J_{0,t}^{a}\\\noalign{\vspace{15pt}}
J_{1,t}^{a}
\end{array}\right) = {\cal D}_{1}^{ab} \left(\begin{array}{c}
\displaystyle\frac{\delta H_{n}}{\delta J_{0}^{b}}\\\noalign{\vspace{6pt}}
\displaystyle\frac{\delta H_{n}}{\delta J_{1}^{b}}
\end{array}\right) = {\cal D}_{2}^{ab} \left(\begin{array}{c}
\displaystyle\frac{\delta \tilde{H}_{n}}{\delta J_{0}^{b}}\\\noalign{\vspace{6pt}}
\displaystyle\frac{\delta \tilde{H}_{n}}{\delta J_{1}^{b}}
\end{array}\right)\;.
\end{equation}

The anti-symmetry of the Hamiltonian structures in
(\ref{hamiltonianstructure}) is manifest and it can be
seen, using the method of prolongation~\cite{prolongation}, that they satisfy Jacobi
identity. Therefore, these are genuine Hamiltonian structures of the
system. Furthermore, it is also easy to check, using the method of
prolongation, that
\begin{equation}
{\cal D}_{1} + \xi\, {\cal D}_{2}\,
\end{equation}
satisfies Jacobi identity, where $\xi$ is an arbitrary constant
parameter. As a result, the system has two compatible Hamiltonian
structures and is truly a bi-Hamiltonian system. However, it is worth
noting that, unlike in conventional integrable systems, here the two
Hamiltonian structures act on different families of charges to yield
the same dynamical equations. Since the criterion of
Magri~\cite{bihamiltonian} applies to
systems where the bi-Hamiltonian structures act on the same family of
conserved charges, it is not {\em a priori} clear that the
integrability of the system follows from the existence of a
bi-Hamiltonian structure. This is a  question that we will
take up in the next section. We would like to point out here that
under the action of ${\cal D}_{1}$, the conserved charge
$\tilde{H}_{1}$ generates spatial translations, as does $H_{1}$ under
the action of ${\cal D}_{2}$. This is another interesting feature of
this system, namely, the ``Hamiltonian'' and the ``momentum'' (with
respect to a given Hamiltonian structure) seem to be distributed into
distinct families of conserved quantities in the non-linear sigma
model.

\section{Complete integrability of the system:}

Before proving complete integrability of the non-linear sigma model,
let us derive the recursion operators that relate the conserved
charges of the two infinite sets. Using the relations in
(\ref{extrarelations}), it is straightforward to show that the Lax
operator, $S^{ab}$, defines a recursion operator for the system. It
can be explicitly checked that
\begin{equation}
\left(\begin{array}{c}
\displaystyle\frac{\delta H_{n+1}}{\delta J_{0}^{a}}\\\noalign{\vspace{6pt}}
\displaystyle\frac{\delta H_{n+1}}{\delta J_{1}^{a}}
\end{array}\right) = S^{ab} \left(\begin{array}{c}
\displaystyle\frac{\delta H_{n}}{\delta J_{0}^{b}}\\\noalign{\vspace{6pt}}
\displaystyle\frac{\delta H_{n}}{\delta J_{1}^{b}}
\end{array}\right),\qquad \left(\begin{array}{c}
\displaystyle\frac{\delta \tilde{H}_{n+1}}{\delta J_{0}^{a}}\\\noalign{\vspace{6pt}}
\displaystyle\frac{\delta \tilde{H}_{n+1}}{\delta J_{1}^{a}}
\end{array}\right) = S^{ab} \left(\begin{array}{c}
\displaystyle\frac{\delta \tilde{H}_{n}}{\delta J_{0}^{b}}\\\noalign{\vspace{6pt}}
\displaystyle\frac{\delta \tilde{H}_{n}}{\delta J_{1}^{b}}
\end{array}\right)\;.
\end{equation}
Namely, the charges within each of the two sets are related
recursively by the Lax operator, $S^{ab}$, itself.

Furthermore, let us define
\begin{equation}
\tilde{S}^{ab} =\left({\cal D}^{-1}_{1}\right)^{ac}\, {\cal D}_{2}^{cb} = \delta^{ab}
\sigma_{1} + \partial^{-1}\left(\begin{array}{cc}
f^{abc} J_{0}^{c} & f^{abc} J_{1}^{c}\\\noalign{\vspace{6pt}}
0 & 0
\end{array}\right)\;.\label{recursion1}
\end{equation}
Using (\ref{extrarelations}), it is easy to check that the second term
on the right hand side of (\ref{recursion1}) gives zero acting on
\begin{equation}
\left(\begin{array}{c}
\displaystyle\frac{\delta H_{n}}{\delta J_{0}^{b}}\\\noalign{\vspace{6pt}}
\displaystyle\frac{\delta H_{n}}{\delta J_{1}^{b}}
\end{array}\right)\;,\qquad {\rm or}\qquad \left(\begin{array}{c}
\displaystyle\frac{\delta \tilde{H}_{n}}{\delta J_{0}^{b}}\\\noalign{\vspace{6pt}}
\displaystyle\frac{\delta \tilde{H}_{n}}{\delta J_{1}^{b}}
\end{array}\right)\;.\label{gradient}
\end{equation}
It follows then that
\begin{equation}
\left(\begin{array}{c}
\displaystyle\frac{\delta \tilde{H}_{n}}{\delta J_{0}^{a}}\\\noalign{\vspace{6pt}}
\displaystyle\frac{\delta \tilde{H}_{n}}{\delta J_{1}^{a}}
\end{array}\right) = \tilde{S}^{ab} \left(\begin{array}{c}
\displaystyle\frac{\delta {H}_{n}}{\delta J_{0}^{b}}\\\noalign{\vspace{6pt}}
\displaystyle\frac{\delta {H}_{n}}{\delta J_{1}^{b}}
\end{array}\right) = \sigma_{1} \left(\begin{array}{c}
\displaystyle\frac{\delta {H}_{n}}{\delta J_{0}^{a}}\\\noalign{\vspace{6pt}}
\displaystyle\frac{\delta {H}_{n}}{\delta J_{1}^{a}}
\end{array}\right)\;.
\end{equation}
This is, in fact,  already noted (the last relation) in
(\ref{extrarelations}). It is worth noting here that the
recursion operator, (\ref{recursion1}), conventionally constructed
from the  two Hamiltonian structures in
an integrable system, in this case relates charges of the two distinct
sets. In fact, on the space of the gradients of the Hamiltonian,
(\ref{gradient}), it is easy to verify that
\begin{equation}
(\tilde{S})^{2} = I\;.
\end{equation}
Namely, although the square of this recursion operator is not identity,
its effect on the space (\ref{gradient}) is that of the identity
operator. This is quite important in that it implies that there are
only two infinite sets of local conserved charges (without
derivatives) associated with the system.

Thus, we see that, in the case of the non-linear sigma model, there
exist two recursion operators --- $S^{ab}$ relates charges within a
family whereas $\tilde{S}^{ab}$ relates charges between the two
families of conserved charges. By construction $\tilde{S}^{ab}$ is
related to the two Hamiltonian structures of the system. One can
similarly ask if there is yet another Hamiltonian structure within a
given family of charges constructed in the conventional manner as
\begin{equation}
\overline{\cal D}^{ab} = {\cal D}_{1}^{ac}\, S^{cb}\;.
\end{equation}
We have checked that this structure does not satisfy Jacobi
identity. (In fact, this structure does not even have the required
anti-symmetry property. If it is anti-symmetrized by hand, the new
structure does not lead to the correct dynamical equations, in
addition to the fact that the Jacobi identity does not hold.) This,
therefore, remains an open question.

To prove complete integrability of the system, we note that
\begin{equation}
\left\{H_{n},H_{m}\right\}_{1} = \int dx\left[\frac{\partial
h_{n}}{\partial J_{0}^{a}}\left(\partial \frac{\partial
h_{m}}{\partial J_{1}^{a}}\right) + \frac{\partial h_{n}}{\partial
J_{1}^{a}}\left(\partial \frac{\partial h_{m}}{\partial
J_{0}^{a}}-f^{abc}J_{0}^{c} \frac{\partial h_{m}}{\partial
J_{1}^{b}}\right)\right]\;.
\end{equation}
From the forms of the conserved charges in (\ref{conserveddensity}),
it is easy to see that the term on the right hand side involving the
structure constant does not contribute. As a result, we have
\begin{eqnarray}
\left\{H_{n},H_{m}\right\}_{1} \equal \int dx\left[\frac{\partial
h_{n}}{\partial J_{0}^{a}}\partial \frac{\partial h_{m}}{\partial
J_{1}^{a}}+ \frac{\partial h_{n}}{\partial J_{1}^{a}}\partial
\frac{\partial h_{m}}{\partial
J_{0}^{a}}\right]\nonumber\\\noalign{\vspace{6pt}}
 \equal \int dx\left[F^{a} J_{0,x}^{a} + G^{a}
J_{1,x}^{a}\right]\;,\label{involution1}
\end{eqnarray}
where
\begin{eqnarray}
F^{a} \equal \frac{\partial h_{n}}{\partial J_{0}^{b}}
\frac{\partial^{2} h_{m}}{\partial J_{0}^{a}\,\partial J_{1}^{b}} +
\frac{\partial h_{n}}{\partial J_{1}^{b}} \frac{\partial^{2}
h_{m}}{\partial J_{0}^{a}\,\partial
J_{0}^{b}}\;,\nonumber\\\noalign{\vspace{6pt}}
G^{a} \equal \frac{\partial h_{n}}{\partial J_{0}^{b}}
\frac{\partial^{2} h_{m}}{\partial J_{1}^{a}\,\partial J_{1}^{b}} +
\frac{\partial h_{n}}{\partial J_{1}^{b}} \frac{\partial^{2}
h_{m}}{\partial J_{1}^{a}\,\partial J_{0}^{b}}\;.
\end{eqnarray}
It is straightforward now to check, using (\ref{extrarelations}), that
\begin{equation}
\frac{\partial F^{a}}{\partial J_{1}^{b}} - \frac{\partial
G^{a}}{\partial J_{0}^{b}} = 0\;.
\end{equation}
This implies that there exists a function $X$ such that
\begin{equation}
F^{a} = \frac{\partial X}{\partial J_{0}^{a}}\;,\qquad G^{a} =
\frac{\partial X}{\partial J_{1}^{a}}\;,\qquad \frac{\partial^{2}
X}{\partial J_{0}^{a}\,\partial J_{1}^{b}} = \frac{\partial^{2}
X}{\partial J_{0}^{b}\,\partial J_{1}^{a}}\;.
\end{equation}
Substituting this into (\ref{involution1}), we obtain
\begin{equation}
\left\{H_{n},H_{m}\right\}_{1} = \int dx\;\frac{dX}{dx} = 0\;,
\end{equation}
with the usual assumptions on the asymptotic fall off of the fields. In a
similar manner, it can be shown that
\begin{equation}
\left\{H_{n},\tilde{H}_{m}\right\}_{1} = 0 =
\left\{\tilde{H}_{n},\tilde{H}_{m}\right\}_{1}\;.
\end{equation}
Namely, all the charges are in involution with respect to the first
Hamiltonian structure. The involution with respect to the second
Hamiltonian structure is also easy to show along similar lines,
\begin{equation}
\left\{H_{n},H_{m}\right\}_{2} = 0 =
\left\{H_{n},\tilde{H}_{m}\right\}_{2} =
\left\{\tilde{H}_{n},\tilde{H}_{m}\right\}_{2}\;.
\end{equation}
This proves that the non-linear sigma model is completely integrable.

\section{Discussions:}

In this letter, we have proposed a Lax equation for the non-linear
sigma model which directly generates all the conserved local
quantities (without any derivative) of the system. There are, in fact,
two infinite sets of conserved charges and the similarities of this
system with the two component Riemann equation are discussed. We have
obtained the two Hamiltonian structures of the system and have shown
that they are compatible thereby proving that the system is
bi-Hamiltonian. However, unlike conventional integrable systems, here
the two Hamiltonian structures act on charges in the different sets to
give  the
same dynamical equations. As a result, integrability of the system
does not follow in spite of the fact that it is a bi-Hamiltonian
system. We have constructed two recursion operators for the system --- one
that relates conserved quantities withing any given family and the
other that relates the charges between the two families. We have
explicitly shown that the two sets of charges are in involution with
respect to either of the Hamiltonian structures, thereby proving the
integrability of the system. We have also brought out several other
interesting features associated with the system.

In the conventional discussion of the non-linear sigma
model~\cite{zerocur,nonlocalcharges}, there
is a standard Hamiltonian structure given by
\begin{equation}
{\cal D}_{\rm std}^{ab} = \left(\begin{array}{cc}
f^{abc}J_{0}^{c} & \delta^{ab} \partial + f^{abc}J_{1}^{c}\\\noalign{\vspace{6pt}}
\delta^{ab} \partial + f^{abc} J_{1}^{c} & 0
\end{array}\right)\;.\label{standard}
\end{equation}
This is manifestly anti-symmetric and can be easily checked to satisfy
Jacobi identity. On the other hand, this seems to be quite different
from the two Hamiltonian structures that we have obtained in
(\ref{hamiltonianstructure}). We note that the Hamiltonian structure
in (\ref{standard}) acting on $H_{1}$ leads to the non-linear sigma
model equations, (\ref{sigmamodelequations}). On the other hand, we
note that
\begin{equation}
{\cal D}_{\rm std}^{ab} - {\cal D}_{1}^{ab} = \left(\begin{array}{cc}
f^{abc}J_{0}^{c} & f^{abc}J_{1}^{c}\\\noalign{\vspace{6pt}}
f^{abc}J_{1}^{c} & f^{abc}J_{0}^{c}
\end{array}\right)\;.
\end{equation}
Using (\ref{extrarelations}), it is easy to see that this difference
gives vanishing contribution on the space (\ref{gradient}). As a
result, on the space (\ref{gradient}), the two structures ${\cal
D}_{\rm std}^{ab}$ and ${\cal D}_{1}^{ab}$ have identical effect. (It
is interesting to note that, in spite of this, the two structures,
${\cal D}_{1}^{ab}$ and ${\cal D}_{\rm std}^{ab}$ are not compatible.)
This feature
is completely new, namely, we have two Hamiltonian structures,
satisfying Jacobi identity, which give the same equations acting on
the same Hamiltonian (normally multi-Hamiltonian structures act on
different Hamiltonians). It is not {\em a priori} clear, therefore,
how one can select between these two structures. The difference in the
two Hamiltonian structures lies in their action on the space of
charges that are not gauge invariant (which have a free internal
index). Namely, we know that the non-linear sigma model has an
infinite set of conserved non-local charges, the first of which is, in
fact, local and is of the form
\begin{equation}
Q^{a} = \int dx\; J_{0}^{a}\;.
\end{equation}
This charge is, in fact, supposed to generate the global rotations of
the system through the action of the Hamiltonian structure of the
system. It can be checked that the action of ${\cal D}_{\rm std}^{ab}$
acting on this leads to the correct rotation while ${\cal D}_{1}^{ab}$
does not. This may be a reason to choose the conventional Hamiltonian
structure over ${\cal D}_{1}^{ab}$. These are, however, open questions
that need further study.

\section*{Acknowledgments}

This work was supported in part by US DOE grant
no. DE-FG-02-91ER40685 as well as by CNPq, Brasil.

\end{document}